\documentclass[11pt]{article}

\usepackage[english]{babel}
\usepackage[T1]{fontenc}
\usepackage[latin1]{inputenc}
\usepackage{amssymb}
\usepackage{amsfonts}
\usepackage{amsmath}
\usepackage{bm}
\usepackage{subfigure}
\usepackage{graphicx}
\usepackage{color}
\usepackage{authblk}

\def\Bmp#1{ \begin{minipage}{#1} }
\def\Bmpc#1{ \begin{minipage}[c]{#1} }
\def\Bmpt#1{ \begin{minipage}[t]{#1} }
\def\Bmpb#1{ \begin{minipage}[b]{#1} }
\def\Emp{ \end{minipage} }

\def\J{{\mathcal{J}}}
\def\E{{\mathcal{E}}}

\def\P{{\mathcal{P}}}
\def\Q{{\mathcal{Q}}}
\def\R{{\mathcal{R}}}

\def\K{{\mathcal{K}}}
\def\Q{{\mathcal{Q}}}

\def\twKP{\tilde{\omega}_{\K_0,\P_0}}

\def\tf0{\tilde{\varphi}_{0}}

\def\0{{\bf 0}}

\newcommand{\xvec}{\mathbf{x}}
\newcommand{\yvec}{\mathbf{y}}
\newcommand{\uvec}{\mathbf{u}}
\newcommand{\kvec}{\mathbf{k}}

\newcommand{\uhat}{\widehat{u}}

\newcommand{\laplacian}{\Delta}
\newcommand{\nablaperp}{\nabla^{\perp}}

\newcommand{\tuvec}{\widetilde{\mathbf{u}}}
\newcommand{\Reyn}{\textrm{Re}}

\newcommand{\tuvecKP}{\widetilde{\mathbf{u}}_{\K_0,\P_0}}
\newcommand{\tuvecRP}{\widetilde{\mathbf{u}}_{\Reyn_0,\P_0}}
\newcommand{\twRP}{\widetilde{\omega}_{\Reyn_0,\P_0}}

\begin{document}

\title{Maximum palinstrophy amplification in the two-dimensional Navier-Stokes equations}

\author[1]{Diego Ayala}
\author[1,2,3]{Charles R. Doering}
\author[4]{Thilo M. Simon}

\affil[1]{Department of Mathematics, University of Michigan, Ann Arbor, MI 48109-1043, USA}
\affil[2]{Center for the Study of Complex Systems, University of Michigan, Ann Arbor, MI 48109-1107, USA}
\affil[3]{Department of Physics, University of Michigan, Ann Arbor, MI 48109-1040, USA}
\affil[4]{Max Planck Institute for Mathematics in the Sciences, Inselstra\ss e 22, 04103 Leipzig, Germany }

\renewcommand\Authands{ and }

\date{\today}

\maketitle

\begin{abstract}
We derive and assess the sharpness of analytic upper bounds for the instantaneous growth rate and finite-time amplification of palinstrophy in solutions of the two-dimensional incompressible Navier-Stokes equations. A family of optimal solenoidal fields parametrized by initial values for the Reynolds number $\Reyn$ and palinstrophy $\P$ which maximize $d\P/dt$ is constructed by numerically solving suitable optimization problems for a wide range of $\Reyn$ and $\P$, providing numerical evidence for the sharpness of the analytic estimate $d\P/dt \leq \left(a + b\sqrt{\ln\Reyn+c} \, \right) \P^{3/2}$ with respect to both $\Reyn$ and $\P$.
This family of instantaneously optimal fields is then used as initial data in fully resolved direct numerical simulations and the time evolution of different relevant norms is carefully monitored as the palinstrophy is transiently amplified before decaying.
The peak values of the palinstrophy produced by these initial data, i.e., $\sup_{t > 0} \P (t)$, are observed to scale with the magnitude of the initial palinstrophy $\P(0)$ in accord with the corresponding \emph{a priori} estimate.
Implications of these findings for the question of finite-time singularity formation in the three-dimensional incompressible Navier-Stokes equation are discussed. 

\end{abstract}

\section{Introduction}
\label{sec:intro}

Energy methods are among the most popular techniques used in the mathematical analysis of evolutionary partial differential equations.
At their core these methods rely on the existence of bounds for a norm $\Q$---an ``energy'' that is not necessarily related to a physical energy---that provides relevant information about the magnitude and regularity (smoothness) of solutions.
The idea is to derive bounds on the growth rate $d\Q/dt$ from the equations of motion utilizing rigorous functional estimates and, given initial data with finite norm $\Q(0)$, to subsequently control $\Q(t)$ for $t>0$ via methods of ordinary differential (in)equations.
In some cases this kind of energy analysis can establish that $\Q(t)$ remains finite---sometimes even uniformly bounded---for all $t>0$ while in others it can only show that that $\Q(t)$ is finite for all time if $\Q(0)$ is sufficiently small.
And in some situations all that can be proved is that $\Q(t) < \infty$ during a finite time interval whose length depends on $\Q(0)$.
Among the many well-known examples amenable to such analysis are the incompressible Navier-Stokes equations.

In the case of unforced flows on the d-dimensional torus ($\mathbb{T}^d$), for example, the $L^2$ norm of the velocity vector field---proportional to the square root of the kinetic energy---decays monotonically due to viscous dissipation so it is bounded uniformly in time by its initial value.
Other norms may grow, however, and their amplification reflects aspects of the cascade processes that characterize much of the complexity of nonlinear fluid mechanics.

In spatial dimension $d = 3$ the enstrophy---the square of the $H^1$ semi-norm of the velocity, which is the same as the $L^2$ norm squared of the vorticity---can be amplified by vortex stretching.
Analysis establishes that the growth rate of enstrophy is a bounded function of the enstrophy, but the resulting differential (in)equations only ensure that the enstrophy remains bounded forever if the initial data is sufficiently small---in particular if the product of the initial kinetic energy and enstrophy is sufficiently small; see, for example, \cite{d09}.
That is, the possibility of finite-time singularities is not ruled out by energy analysis.
The enstrophy corresponds to a particularly relevant norm in this problem because limits on the $H^1$ semi-norm of the velocity can be ``bootstrapped'' to bounds on norms of higher derivatives establishing infinite differentiability of the solutions while the enstrophy remains finite.
On the other hand divergent upper bounds on enstrophy do not guarantee divergent solutions and no ``blow-up'' has been demonstrated to date.
Hence the question of long-time existence of smooth solutions of the three-dimensional (3-D) incompressible Navier-Stokes equations remains one of the grand challenges for mathematical physics.

Much more is known about solutions of the two-dimensional (2-D) Navier-Stokes equations.
In the case of unforced flows on $\mathbb{T}^2$ both the $L^2$ norm and the $H^1$ semi-norm of solutions (i.e., both the kinetic energy and the enstrophy) decay monotonically in time while the $H^2$ semi-norm of the velocity field---which is the same $H^1$ semi-norm of the pseudo-scalar vorticity field, the square of which is known as the {\it palinstrophy}---can be amplified by a vorticity gradient stretching mechanism.
Energy methods can be used to show that the palinstrophy of solutions of 2-D incompressible Navier-Stokes equations remains finite for all time and there are no potential finite-time singularities.
Moreover, energy methods can be used to derive rigorous upper limits on the peak palinstrophy as a function of the norms of the initial data.

In this paper we investigate the quantitative accuracy of palinstrophy amplification bounds in order to evaluate the practical predictive power of energy method analysis.
We consider the incompressible 2-D Navier-Stokes equation, written here as an evolution equation for the vorticity $\omega$, on spatial domain $\Omega = \mathbb{T}^2$ (the $L \times L$ square with periodic boundary conditions):
\begin{equation}\label{eq:2DNSE}
\omega_t + \uvec\cdot\nabla\omega = \nu\laplacian\omega, \quad
\nabla\cdot\uvec = 0, \quad \text{and} \quad \omega(\cdot,0) = \omega_0
\end{equation}
where $\uvec$ is the velocity field, $\nu$ is the kinematic viscosity, and the velocity and the vorticity are related via
\begin{equation*}
\omega = {\hat {\bf e}}_3 \cdot \nabla \times \uvec \ \  \Leftrightarrow \ \ -\laplacian\uvec = \nablaperp\omega
\end{equation*}
with $\nablaperp = \left[\partial_{x_2},-\partial_{x_1}\right]$.
We are interested in the time evolution of the the energy, enstrophy and palinstrophy defined here, respectively, as
\begin{subequations}\label{eq:Q_defn}
\begin{align}
\K\{\uvec(\cdot,t)\} & = \frac{1}{2}\int_\Omega | \uvec(\cdot,t) |^2 \;d\Omega, \\
\E\{\uvec(\cdot,t)\} & = \frac{1}{2}\int_\Omega | \nabla\uvec(\cdot,t) |^2 \;d\Omega = \frac{1}{2}\int_\Omega | \omega(\cdot,t) |^2 \;d\Omega,\\
\P\{\uvec(\cdot,t)\} & = \frac{1}{2}\int_\Omega | \laplacian\uvec(\cdot,t) |^2\;d\Omega = \frac{1}{2}\int_\Omega | \nablaperp\omega(\cdot,t) |^2 \;d\Omega = \frac{1}{2}\int_\Omega | \nabla\omega(\cdot,t) |^2 \;d\Omega.
\end{align}
\end{subequations}    
The temporal evolution of of $\K$, $\E$ and $\P$ are given by
\begin{subequations}\label{eq:dQdt_defn}
\begin{align}
\frac{d\K}{dt} & = -\nu\int_\Omega | \omega(\cdot,t) |^2 \;d\Omega = -2\nu\E\{\uvec(\cdot,t)\}, \label{eq:dKdt_defn} \\
\frac{d\E}{dt} & = -\nu\int_\Omega | \nabla\omega(\cdot,t) |^2 \;d\Omega = -2\nu\P\{\uvec(\cdot,t)\}, \label{eq:dEdt_defn}\\
\frac{d\P}{dt} & = -\nu\int_\Omega | \laplacian\omega(\cdot,t) |^2\;d\Omega - 
\int_{\Omega}\nabla\omega\cdot\nabla\uvec\cdot\nabla\omega \;d\Omega =: \R\{\uvec(\cdot,t)\}, \label{eq:dPdt_defn}
\end{align}
\end{subequations}    
defining, in the last line, the palinstrophy generation rate functional $\R\{\uvec(\cdot,t)\}$.

We are focused on the question of the sharpness of rigorous analytic bounds on the instantaneous and finite-time growth of palinstrophy, the only quantity from \eqref{eq:Q_defn} with nonmonotonic temporal dynamics.
This is of interest because the functional analysis techniques used to derive the estimates rely only on the structure of the palinstrophy generation rate functional and fundamental relations between norms and {\it not} specifically on the physics associated to the problem at hand, making this study relevant to other partial differential equations.

Dimensional analysis requires that bounds on the instantaneous rate of palinstrophy generation, $\R\{\uvec(\cdot,t)\} = d\P/dt$ as a function of the instantaneous energy, enstrophy, palinstrophy, viscosity $\nu$ and domain length scale $L$---if such bounds exist at all---must be of the form
\begin{equation}\label{eq:gen_Estimate}
\R\{\uvec(\cdot,t)\} \leq \Gamma(\K,\E,\P,\nu,L) \ \P^{3/2}
\end{equation} 
where the prefactor $\Gamma$ is a dimensionless function of dimensionless combinations of the energy, enstrophy, palinstrophy, viscosity and system size.
An estimate of this form \eqref{eq:gen_Estimate} will be declared sharp if and only if
\begin{equation*}
\max_{\uvec \in \mathcal{S}} \P\{\uvec\}^{-3/2} \R\{\uvec\} \sim \Gamma(\K,\E,\P,\nu,L)
\end{equation*}
where the maximum is over the set $\mathcal{S}$ of all spatially periodic divergence-free vector fields with energy, enstrophy and palinstrophy values $\K$, $\E$ and $\P$ on the  $L \times L$ square. Thus the sharpness of such estimates can be addressed by solving the constrained optimization problem
\begin{equation}
\tuvec_{\mathcal{S}} = \mathop{\arg\max}_{\uvec \in \mathcal{S}} \; \R\{\uvec\}
\end{equation}
where the constraint manifold $\mathcal{S}$ can be interpreted as the intersection (in infinite-dimensional space) of spheres of given radius measured in different norms.

Given the anisotropic nature of the constraint manifold it is desirable to seek bounds and test estimates with the fewest possible number of parameters in the prefactor $\Gamma$.
Moreover, if we are interested in estimates that could conceivably make sense in the infinite volume limit, i.e., if we seek optimizing flow structures and bounds that are {\it independent} of the domain scale $L$, then the prefactor can only be a function of the dimensionless combinations $\K^{1/2}/\nu$---a ratio that is naturally referred to as the Reynolds number $\Reyn$---and $(\K \P)^{1/2} / \E$.
If we further conjecture an $L$-independent estimate for $\R\{\uvec(\cdot,t)\}$ whose large palinstrophy behavior is dominated non-trivally by the leading $\P^{3/2}$ in (\ref{eq:gen_Estimate}) then the asymptotic prefactor can be a function of  $\Reyn$ alone.
That is, the prefactor would be of the form $\Gamma(\K,\E,\P,\nu,L) = \gamma(\Reyn)$.
In fact there are rigorous upper bounds on the palinstrophy generation rate of the form $ \R \le \gamma(\Reyn) \, \P^{3/2}$.
\cite{ap13a} proved that $\R \lesssim \Reyn \, \P^{3/2}$ and computed optimal vector fields indicating that the $3/2$ exponent for $\P$ is sharp but that the ${\cal O}(\Reyn)$ prefactor is {\it not} sharp.

In this work we assess the sharpness of the improved estimate
\begin{equation}\label{eq:dPdt_LogEstimate}
\begin{aligned}
\frac{d\P}{dt} & = \R\{\uvec\} \le \left(a + b\sqrt{\ln\Reyn+c}  \right) \P^{3/2} \\ 
\text{with} \quad & a = 0, \quad  b = \sqrt{2\pi}, \quad c = -\ln\left( \frac{2}{\sqrt{\pi}} \right)
\end{aligned}
\end{equation} 
for sufficiently large $\Reyn$ and $\P$.
It is derived in Appendix \ref{sec:dPdt_Estimate}.
We demonstrate (1) the sharpness of the upper bound (\ref{eq:dPdt_LogEstimate}) and (2) the extent to which the Navier-Stokes flow starting from instantaneously optimal fields $\tuvec_{\mathcal{S}}$ saturates the corresponding finite-time estimates for palinstrophy amplification
\begin{equation}
\P^{1/2}(t) - \P^{1/2}(0) \leq \phi(\Reyn_0)  \left[\E(0) - \E(t)\right]
\end{equation} 
and
\begin{equation}
\mathop{\max}_{t > 0} \P(t) \le \psi(\Reyn_0) \, \P(0).
\end{equation} 
That the prefactors $\phi$ and $\psi$ depend on $\uvec$ and $\nu$ only via the (initial) Reynolds number $\Reyn_0 = \K^{1/2}(0)/\nu$ implies in particular that the peak palinstrophy amplification factor depends on the viscosity only through its appearance in the explicit function $\psi(\Reyn_0)$.
We note, however, that our study does {\it not} indicate that our estimate for $\psi(\Reyn)$ is sharp with respect to its Reynolds number dependence.

The remainder of this manuscript is as follows. The instantaneous growth of palinstrophy and structure of the optimal vector fields is reviewed in section \ref{sec:2D_InstOpt}. Finite-time growth and maximal amplification of palinstrophy is studied in section \ref{sec:TimeEvol}. Section \ref{sec:Conclusion} contains a discussion of the results along with conclusion and closing remarks. Detailed derivation (from elementary first principles) of the improved analytic estimate on the (\ref{eq:dPdt_LogEstimate}) can be found in the appendix. 

\section{Instantaneously Optimal Growth of Palinstrophy} 
\label{sec:2D_InstOpt}

Given the dependence of analytic estimate \eqref{eq:dPdt_LogEstimate} on palinstrophy $\P$ and Reynolds number $\textrm{Re} = \K^{1/2}/\nu$, the sharpness of the estimate is addressed by numerically solving the constrained optimization problem for the objective functional $\R$ defined in \eqref{eq:dQdt_defn}:
\begin{equation}\label{eq:OptProb_K0P0}
\qquad\mathop{\max}_{\uvec\in\mathcal{S}_{\K_0,\P_0}} \;\; \R(\uvec)
\end{equation}
where
\begin{equation}
\mathcal{S}_{\K_0,\P_0} = \{ \uvec\in H^3(\Omega) : \nabla\cdot\uvec = 0, \;  \K(\uvec) = \K_0, \; \P(\uvec) = \P_0 \}.
\end{equation}
In order to investigate the asymptotic behavior of $\R$ as $\P_0\to\infty$, \eqref{eq:OptProb_K0P0} is solved numerically for a wide range of values of the energy $\K_0 \in [1,100]$ and some choices of $\P_0 \gg \K_0/C_P^2$, where $C_P = 1/(2\pi)^2$ is the Poincar\'e constant for the unit two dimensional torus. We compute with the numerical value $\nu = 10^{-3}$ for the kinematic viscosity, kept constant in all computations, allowing us to probe the dependence of the optimal rate of growth of palinstrophy for values of the Reynolds number in the range $\Reyn_0 \in [10^3,10^4]$. For a given value of $\Reyn_0$, the value of $\P_0$ defining the constraint manifold $\mathcal{S}_{\K_0,\P_0}$ is chosen so that the optimal vorticity field $\twRP$ of the optimal vector field $\tuvecRP = \arg\max \R(\uvec)$ is sufficiently localized in the computational domain, allowing for the effect of boundaries to be neglected.
A family of optimal fields parametrized by their palinstrophy is then constructed by rescaling the optimal field $\twRP$ using the self-similar approach described in Appendix \ref{sec:SelfSimilarAnalysis}.
Details of the numerical methods for the variational problem \eqref{eq:OptProb_K0P0} can be found in the work of \cite{ap13a}. 

\begin{figure}
\linespread{1.1}
\setcounter{subfigure}{0}
\begin{center}
\subfigure[]{\includegraphics[width=0.75\textwidth]{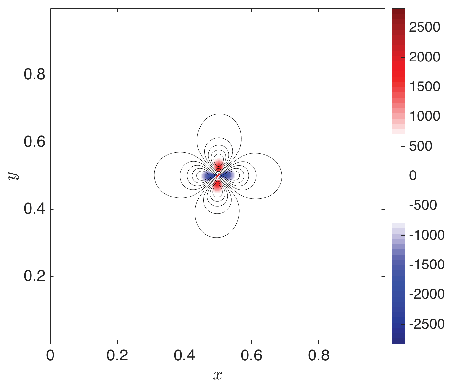}}\\
\subfigure[]{\includegraphics[width=0.75\textwidth]{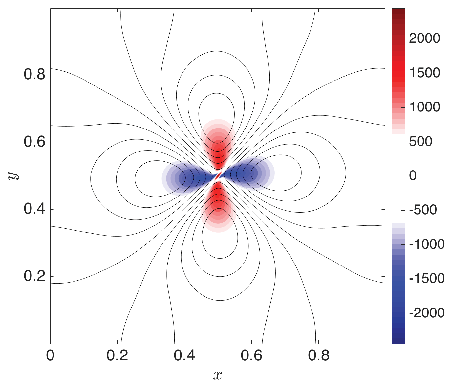}}
\caption[Optimal vorticity field vs $\Reyn$]{Optimal vorticity field $\twRP$ corresponding to $\P_0 = 10^8$ and values of Reynolds number (a) $\Reyn_0 = 10^3$,  and (b) $\Reyn_0 = 10^4$. Streamlines corresponding to selected level sets of the stream function $\psi = -\laplacian^{-1}\twRP$ are shown for reference.
}
\label{fig:OptimVort_vsK_P07}
\end{center}
\end{figure}

Figures \ref{fig:OptimVort_vsK_P07}(a) and \ref{fig:OptimVort_vsK_P07}(b) show the optimal vorticity fields $\twRP$ corresponding to values of Reynolds number $\Reyn_0 = 10^3$ and $\Reyn_0 = 10^4$, respectively, for a fixed value of palinstrophy $\P_0 = 10^8$.  In each case, the optimal vorticity field consists of a vortex quadrupole of finite area, and a localized region of strong vorticity responsible for the extreme growth of palinstrophy. The size of the optimal vortical structure, relative to the size of the domain $\Omega$, has a positive correlation with the Reynolds number. A thorough discussion of the properties of the optimal vorticity fields, and their corresponding time evolution under 2-D Navier-Stokes dynamics can be found in \cite{ap13a} and \cite{ap13b}.

\begin{figure}
\linespread{1.1}
\begin{center}
\includegraphics[width=0.75\textwidth]{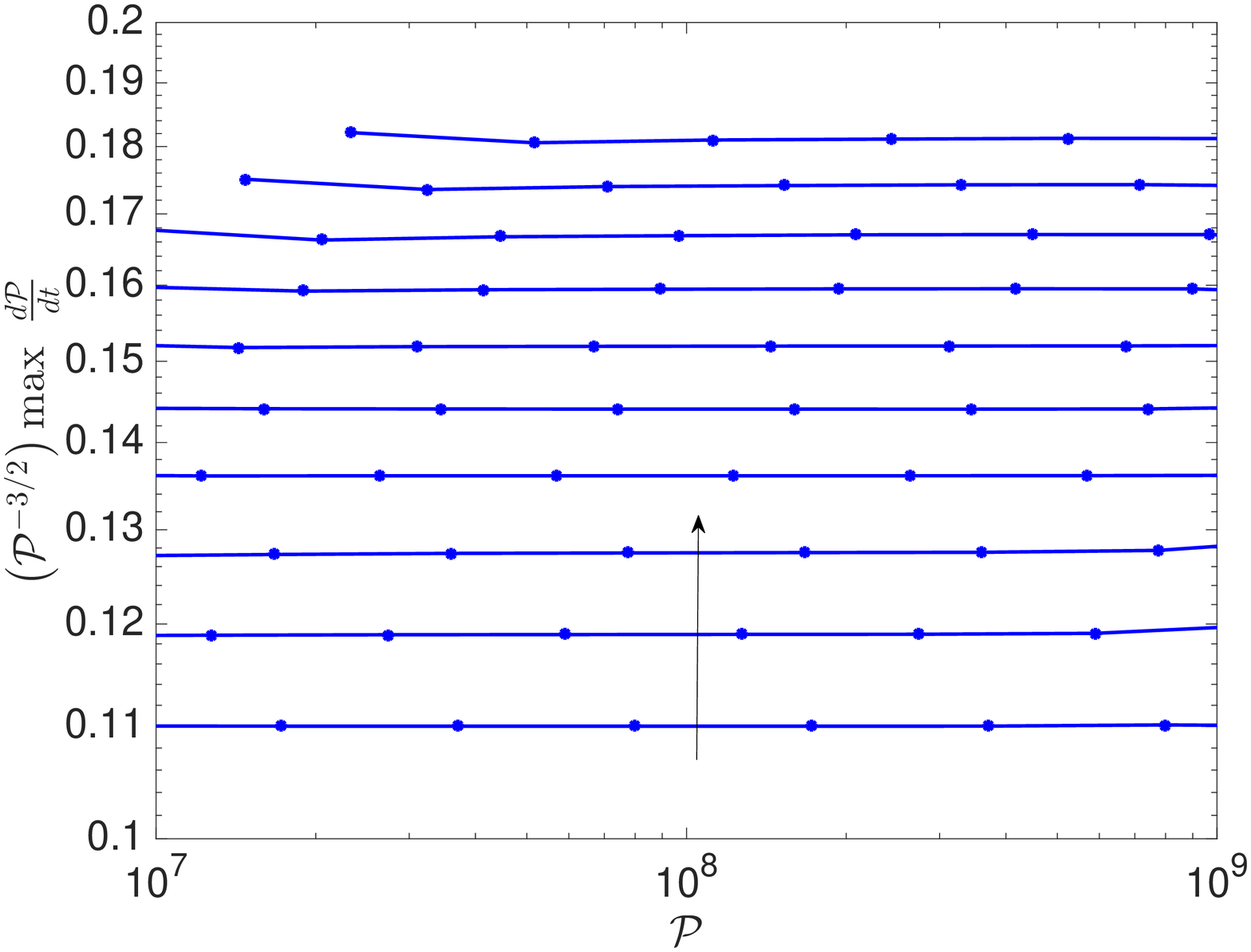}
\caption[Optimal $d\P/dt$ vs $\P$]{
Compensated optimal instantaneous rate of growth of palinstrophy $\widetilde{\R}_{\Reyn_0,\P_0} = \P^{-3/2}_0\R(\tuvecKP)$ as a function of $\P_0$, for values of Reynolds number $\Reyn_0\in[10^3,10^4]$.  The arrow indicates the direction of increasing $\Reyn_0$. 
}
\label{fig:maxdP_vsP}
\end{center}
\end{figure}

To verify the power-law behavior of the optimal instantaneous rate of production of palinstrophy and assess the sharpness of estimate \ref{eq:dPdt_LogEstimate} with respect to the exponent $\alpha = 3/2$, figure \ref{fig:maxdP_vsP} shows the dependence on $\P_0$ of the compensated optimal rate of growth of palinstrophy 
\begin{equation}
\widetilde{\R}_{\Reyn_0,\P_0} = \P^{-3/2}_0\;\R(\tuvecRP),
\end{equation}
for different values of Reynolds number in the interval $\Reyn_0\in [10^3,10^4]$.  The data shown here is an extension of the results reported by \cite{ap13a}. To streamline our discussion, we only include the portion of the data where a clear power-law behavior for 
$\R(\tuvecRP)$ is observed, corresponding to values of palinstrophy much larger than the Poincar\'e limit $\P_0 \to (\nu\Reyn_0/C_P)^2$. As expected from the fact that estimate \eqref{eq:dPdt_LogEstimate} is sharp with respect to the exponent $\alpha = 3/2$, figure \ref{fig:maxdP_vsP} shows that the compensated optimal rate of growth of palinstrophy $\widetilde{\R}_{\Reyn_0,\P_0}$, which corresponds to the prefactor $\gamma(\Reyn)$, is indeed independent of palinstrophy in the limit $\P_0\to\infty$. 

\begin{figure}
\linespread{1.1}
\setcounter{subfigure}{0}
\begin{center}
\subfigure[]{\includegraphics[width=0.75\textwidth]{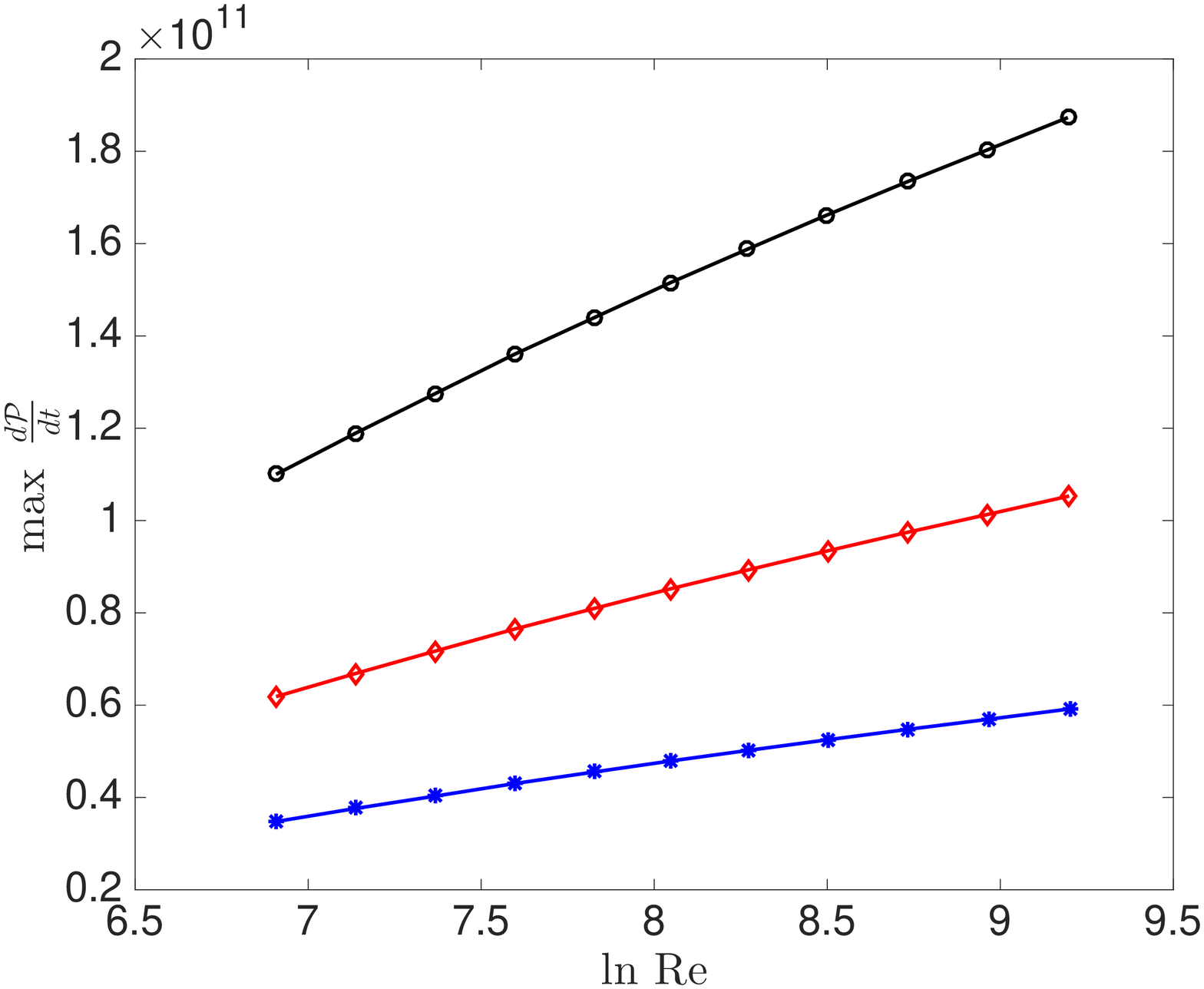}}\\
\subfigure[]{\includegraphics[width=0.75\textwidth]{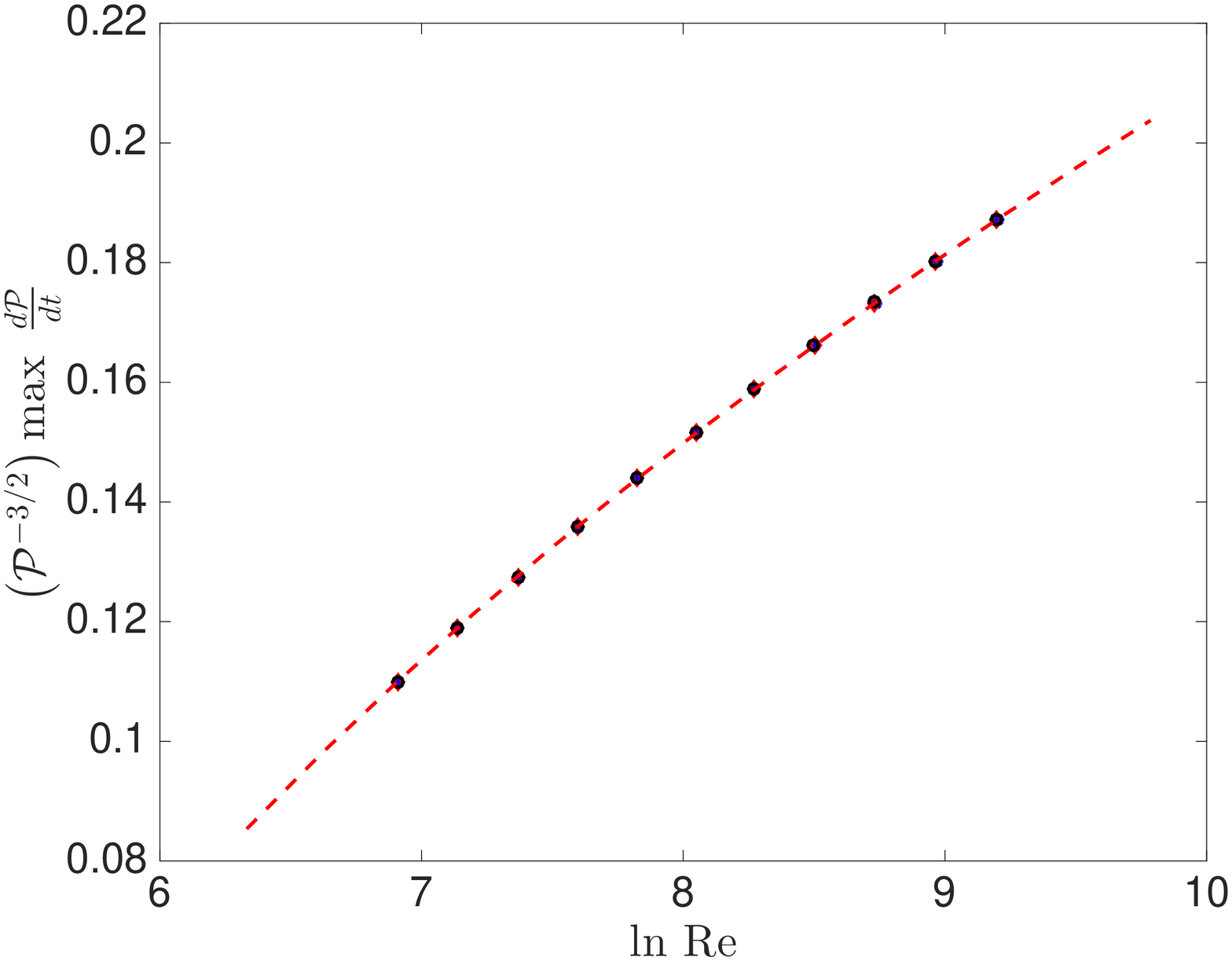}}
\caption[Optimal $d\P/dt$ vs $\Reyn$]{
(a) Optimal rate of growth of palinstrophy, $\R(\tuvecKP)$, as a function of $\ln \Reyn_0$, 
for values of palinstrophy $\P_0 = 4.6\times 10^7$ (blue stars), $\P_0 = 6.8\times 10^7$ (red diamonds) and $\P_0 = 10^8$ (black circles). (b) Compensated optimal rate of growth of palinstrophy, $\P_0^{-3/2}\R(\tuvecKP)$, as a function of $\ln \Reyn_0$, for the same values of palinstrophy as in (a). All curves collapse onto a single ``universal'' curve of the form $\tilde{a} + \tilde{b}\sqrt{\ln \Reyn_0 + \tilde{c} }$ with $\tilde{a} = -0.093$, $\tilde{b} = 0.128$ and $\tilde{c} = -4.38$, shown as a red dashed curve.}
\label{fig:maxdP_vsRe}
\end{center}
\end{figure}

On the other hand, to assess the sharpness of estimate \eqref{eq:dPdt_LogEstimate} with respect to the prefactor 
\[
\gamma(\Reyn) = a + b\sqrt{\ln\Reyn + c},
\]
figures \ref{fig:maxdP_vsRe}(a) and \ref{fig:maxdP_vsRe}(b) show the dependence of the optimal rate of growth of palinstrophy $\R(\tuvecRP)$ on $\ln \Reyn_0$ and of the compensated optimal rate of growth of palinstrophy $\widetilde{\R}_{\Reyn_0,\P_0}$ on $\ln \Reyn_0$, respectively, for three different values of palinstrophy $\P_0 = 4.6\times10^7$, $\P_0 = 6.8\times 10^7$ and $\P_0 = 10^8$.
It can be observed in figure \ref{fig:maxdP_vsRe}(b) that all data points collapse into a single curve of the form
\begin{equation}\label{eq:Prefactor_Fitted}
\gamma_{\Reyn_0} = \tilde{a} + \tilde{b}\sqrt{ \ln\Reyn_0 + \tilde{c} }
\end{equation}
with fitted parameters
\begin{equation}
\quad \tilde{a} = -0.093, \quad \tilde{b} = 0.128, \quad \tilde{c} = -4.38
\end{equation}
shown in the figure as a red dashed curve.
The values of $\tilde{a}$, $\tilde{b}$ and $\tilde{c}$ are obtained by averaging over $\P_0$ the values of $a_{\P_0}$, $b_{\P_0}$ and $c_{\P_0}$ corresponding to the parameters that provide the least-squares fit of the data to a model of the same form as estimate \eqref{eq:dPdt_LogEstimate}.
Although the values of the fitted parameters $\tilde{a}$, $\tilde{b}$ and $\tilde{c}$ differ from the corresponding values in estimate \eqref{eq:dPdt_LogEstimate}, the fundamental dependence of $\gamma(\Reyn)$ on $\Reyn$ is correctly captured by the behavior of 
$\gamma(\Reyn_0) = \widetilde{\R}_{\Reyn_0,\P_0}$ providing positive evidence for the sharpness of estimate \eqref{eq:dPdt_LogEstimate} with respect to the prefactor $\gamma(\Reyn)$.
To summarize, the information presented in Figure \ref{fig:maxdP_vsP} and Figures \ref{fig:maxdP_vsRe}(a)-(b) indicates that the estimate
\[
\frac{d\P}{dt} \leq \left( a + b\sqrt{ \ln\Reyn +c} \right)\P^{3/2} = 
\gamma(\Reyn)\,\P^{3/2}
\]
is indeed sharp with respect to \emph{both} the exponent $\alpha = 3/2$ and the functional form of the prefactor $\gamma(\Reyn) = a + b\sqrt{\ln \Reyn + c}$.

\begin{figure}
\linespread{1.1}
\begin{center}
\includegraphics[width=0.75\textwidth]{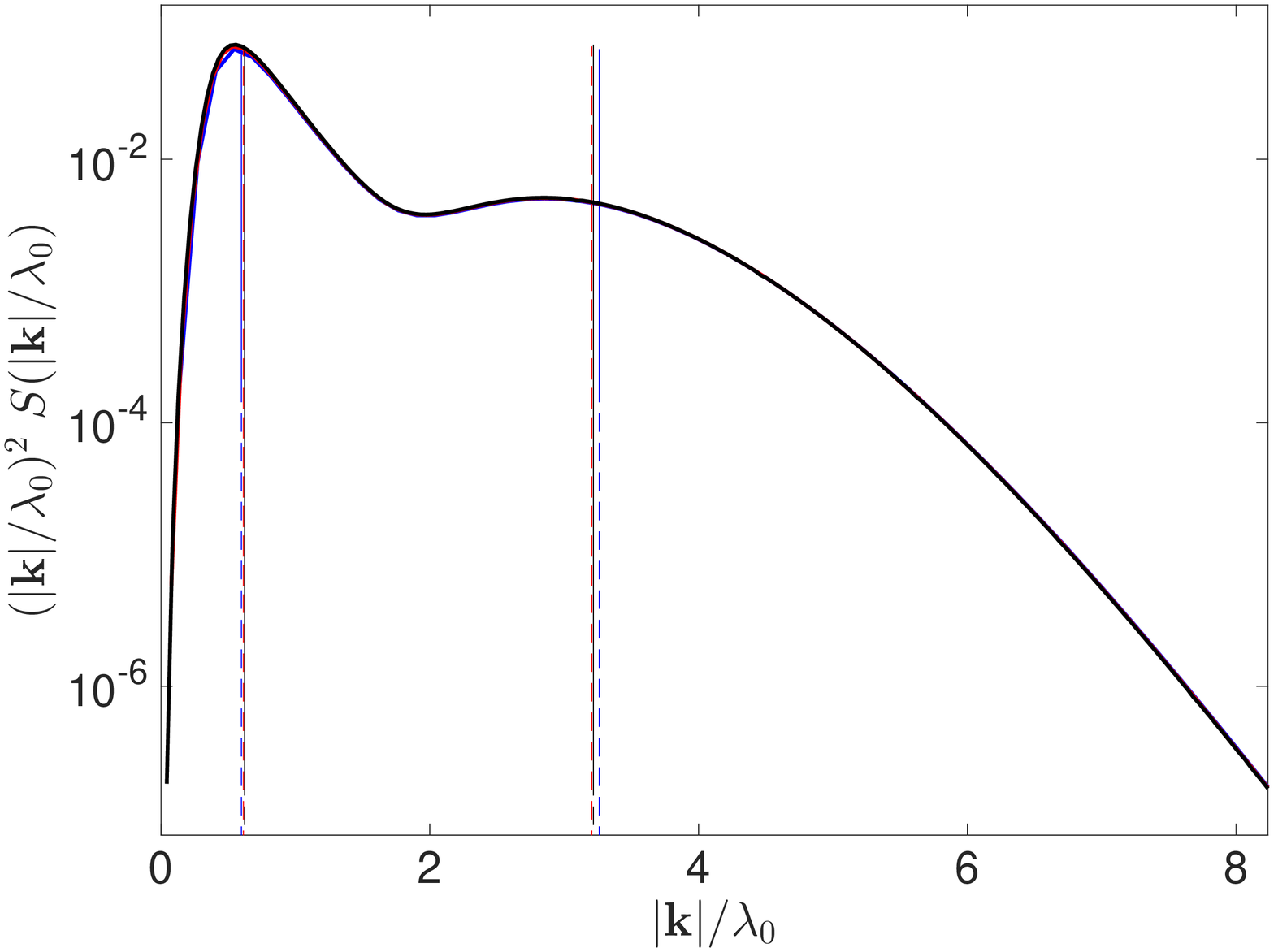}
\caption[Spectral density]{Rescaled compensated spectral density $\left(|\kvec| / \lambda_0 \right)^2 \, S (|\kvec| / \lambda_0)$, for $\lambda_0$ defined in equation \eqref{eq:Lambda_0}, corresponding to $\Reyn = 10^3$ and palinstrophy values $\P_0 = 1.71\times 10^6$, $\P_0 = 1.71\times 10^7$ and $\P_0 = 1.71\times 10^8$. All curves collapse onto a ``universal" spectral density. }
\label{fig:Sk_vsP}
\end{center}
\end{figure}

To complete our analysis of the optimal instantaneous growth of palinstrophy we now look at the structure of the spectrum of the optimal fields $\tuvecRP$. 
A key step in the derivation of estimate \eqref{eq:dPdt_LogEstimate} (see Appendix \ref{sec:dPdt_Estimate}) is the choice of cut-off wave numbers $\Lambda_1$ and $\Lambda_2$ that define the sets $\{ |\kvec| \leq \Lambda_1 \}$, $\{\Lambda_1 < |\kvec| \leq \Lambda_2 \}$ and $\{ |\kvec| > \Lambda_2 \}$ in wavenumber space where $\sum | \kvec | | \uhat(\kvec) |$ displays different behavior.
As discussed in the appendix, the cut-off wave numbers depend on  $\K$, $\P$ and $\nu$ as
\begin{equation*}
\Lambda_1^2 = c_1\frac{\P^{1/2}}{\K^{1/2}}\quad\mbox{and}\quad 
\Lambda_2^2 = \frac{1}{c_2}\frac{\P^{1/2}}{\nu},
\end{equation*}
where $c_1$ and $c_2$ are dimensionless parameters.
Figure \ref{fig:Sk_vsP} shows the rescaled compensated spectral density $(|\kvec|/\lambda_0)^2 \, S(|\kvec|/\lambda_0)$ corresponding to the optimal fields $\tuvecRP$ for $\Reyn = 10^3$ and palinstrophy values $\P_0 = 1.71\times 10^6$, $\P_0 = 1.71\times 10^7$ and $\P_0 = 1.71\times 10^8$. The spectral density $S( |\kvec| )$ is computed as
\[
S(|\kvec|) = \sum_{2\pi k \leq | \kvec | \leq 2\pi(k+1)} |\kvec| |\widehat{\uvec}(\kvec)|^2,
\]
and the scaling factor $\lambda_0$ is given by
\begin{equation}\label{eq:Lambda_0}
\lambda_0 = \frac{\int_0^\infty |\kvec|^2 \; S( |\kvec| ) \; d |\kvec| }
{\int_0^\infty S(|\kvec|)\;d|\kvec|}.
\end{equation}
The wave number $\lambda_1$, computed as
\begin{equation}\label{eq:Lambda_1}
\lambda_1 = \frac{\int_0^\infty |\kvec| \; S( |\kvec| ) \; d |\kvec| }
{\int_0^\infty S(|\kvec|)\;d|\kvec|}
\end{equation}
and the wave number $\lambda_2$ defined as the solution to the equation
\begin{equation}\label{eq:Lambda_2}
\int_0^{\lambda_2} |\kvec|^2 \; S(|\kvec|) \; d|\kvec| = \frac{1}{2}\int_0^{\infty} |\kvec|^2 \; S(|\kvec|) \; d|\kvec| 
\end{equation}
are shown in figure \ref{fig:Sk_vsP} as vertical lines.
These wave numbers correspond to the location of local maxima of the compensated spectral density $|\kvec|^2 \,  S(|\kvec|)$.
As expected from the self-similar construction of the optimal fields $\tuvecRP$, the rescaled compensated spectral densities collapse into a single ``universal'' spectral density, confirming the scale-invariant nature of $\tuvecRP$. 

\section{Finite-time Growth of Palinstrophy}
\label{sec:TimeEvol}

We now focus our attention on the growth of palinstrophy $\P$ over finite time. Time integration of the sharp instantaneous estimate \eqref{eq:dPdt_LogEstimate} and neglect of the time dependence of $\Reyn$ leads to the finite-time estimates 
\begin{equation}\label{eq:maxP_timeDep}
\P^{1/2}(t) - \P^{1/2}(0)  \leq \left( a + b\sqrt{\ln\Reyn_0 + c} \right)\left(\frac{\E(0) - \E(t)}{4\nu}\right)
\end{equation}
and
\begin{equation}\label{eq:maxP_aPriori}
\mathop{\max}_{t>0} \P(t) \leq \psi(\Reyn_0)\P(0), \quad\mbox{with}\quad
\psi(\Reyn_0) = \left(1+ \frac{a + b\sqrt{\ln\Reyn_0 + c}}{4} \,\Reyn_0\right)^2,
\end{equation}
where the prefactor $\psi(\Reyn_0)$ depends exclusively on the initial Reynolds number $\Reyn_0 = \K^{1/2}(0)/\nu$, and the values of $a$, $b$ and $c$ are given in \eqref{eq:Prefactor_Fitted} and obtained by the fitting procedure described in section \ref{sec:2D_InstOpt}. Note that although estimate \eqref{eq:maxP_timeDep} is time-dependent and not in the form power law, it can still be used to determine to what extent the fields saturating a \emph{sharp} instantaneous estimate produce time-dependent flows which also saturate a time-dependent estimate. Also note that while \eqref{eq:maxP_aPriori} is an \emph{a priori} estimate in the form of a power law  which has been obtained from a \emph{sharp} instantaneous estimate, there is no guarantee that it will be sharp with respect to either the exponent or the prefactor in the power law. The derivation of estimates \eqref{eq:maxP_timeDep} and \eqref{eq:maxP_aPriori} can be found in Appendix \ref{sec:maxP_Estimate}.

In order to assess the sharpness of the finite-time estimates \eqref{eq:maxP_timeDep} and \eqref{eq:maxP_aPriori}, we numerically solve the Cauchy problem \eqref{eq:2DNSE} using the instantaneously optimal vorticity fields $\twRP$ as initial condition, and carefully monitor the time evolution of different diagnostics, e.g. $\K(t)$, $\E(t)$ and $\P(t)$, for sufficiently long times. Time integration is performed using an adaptive Runge-Kutta scheme of order 4, and spatial discretization is performed using a pseudo-spectral Fourier-Galerkin method, with the standard ``2/3'' dealiasing rule. The resolution is increased accordingly, ranging from $512^2$ for low-$\Reyn_0$, low-$\P_0$ simulations, to $4096^2$ for high-$\Reyn_0$, 
high-$\P_0$ simulations. We refer the reader to the work by \cite{ap13b} for a thorough discussion of the time evolution of the instantaneously optimal vorticity $\twRP$. 

\begin{figure}
\linespread{1.1}
\begin{center}
\includegraphics[width=0.75\textwidth]{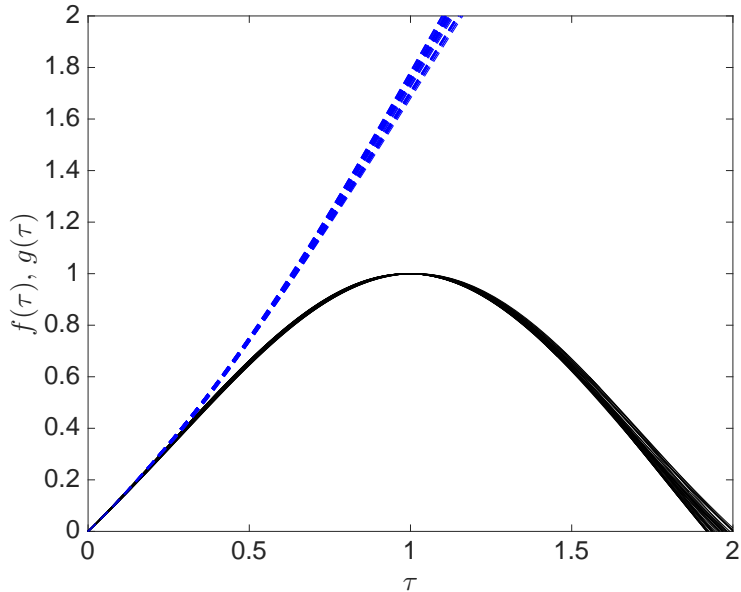}
\caption[Diagnostics $f(\tau)$ and $g(\tau)$]{
Rescaled diagnostics $f(\tau)$ (black solid lines) and $g(\tau)$ (blue dashed lines) for different values of $\P_0$ and $\Reyn_0$. All data collapses onto a ``universal'' pair of curves.} 
\label{fig:maxP_FandG}
\end{center}
\end{figure}

The sharpness of the point-wise estimate \eqref{eq:maxP_timeDep} can be studied by considering the functions:
\[
\varphi (t) = \P^{1/2}(t) - \P^{1/2}(0), \quad \mbox{and} \quad
\mu (t) = \left(\frac{a + b \sqrt{\ln \Reyn_0+c}}{4\nu} \right)\left( \E(0) - \E(t) \right)
\]
and comparing their behavior as functions of time. For this, consider the characteristic time scale $t_{\max}$ and the characteristic palinstrophy scale $\rho$ defined as:
\[
t_{\max} = \mathop{\arg\max}_{t > 0}\, \varphi (t) \quad \mbox{and} \quad
\rho = \mathop{\max}_{t \geq 0}\, \varphi (t),
\]
and the rescaled diagnostics:
\begin{equation}
f(\tau) = \varphi(t_{\max}  \,  \tau)/\rho
\end{equation}
and
\begin{equation}
g(\tau) = \mu(t_{\max} \, \tau)/\rho.
\end{equation}
With these definitions, the point-wise estimate \eqref{eq:maxP_timeDep} simply reads $f(\tau) \leq g(\tau)$. Figure \ref{fig:maxP_FandG} shows the dependence of $f$ and $g$ on the rescaled time $\tau = t/t_{\max}$, corresponding to different values of $\Reyn\in[10^3,10^4]$ and $\P_0 \in [1.7\times 10^6, 1.7\times10^9]$. As expected from the fact that the initial condition is constructed in a self-similar manner, all data collapses onto single ``universal'' curves for $f$ and $g$. The data indicates that estimate \eqref{eq:maxP_timeDep} is saturated only over a short interval, as expected from the fact that the fields are optimal only in an instantaneous sense, and the optimal growth of palinstrophy can be sustained only over this short interval.

\begin{figure}
\linespread{1.1}
\setcounter{subfigure}{0}
\begin{center}
\subfigure[]{\includegraphics[width=0.75\textwidth]{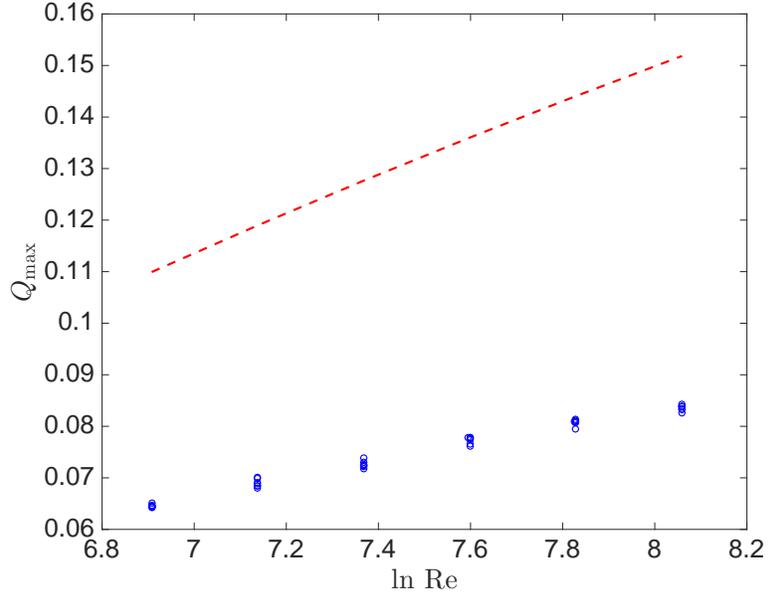}}\\
\subfigure[]{\includegraphics[width=0.75\textwidth]{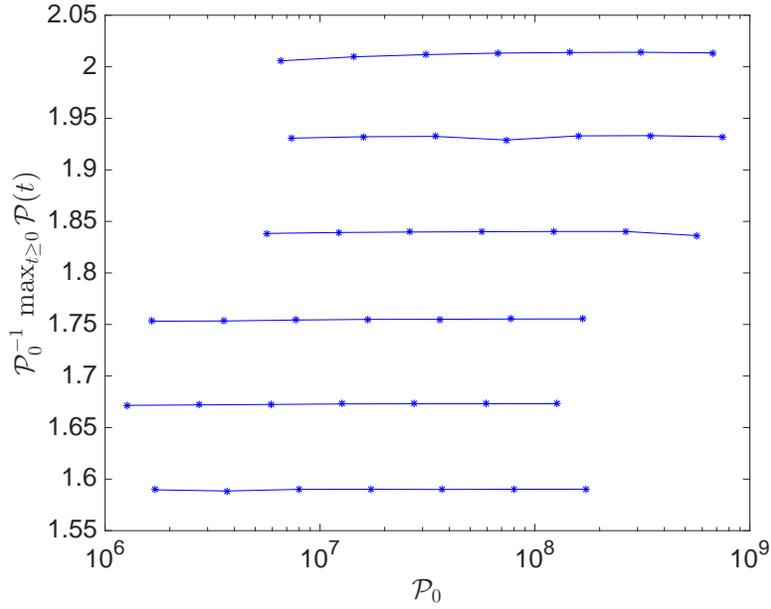}}
\caption[Optimal $P(t)$ vs $\P$]{
(a) The quantity $Q_{\max}$, defined in \eqref{eq:Qmax} as a function of $\ln \Reyn_0$. The prefactor $C_{\Reyn_0} = \tilde{a}+\tilde{b}\sqrt{\tilde{c}+\ln\Reyn_0}$ with $\tilde{a}$, $\tilde{b}$ and $\tilde{c}$ from \eqref{eq:Prefactor_Fitted} is shown as a red dashed line. (b) Compensated maximum palinstrophy, $\P_{\max} = \P_0^{-1}\max_{t > 0}\P(t)$, as a function of  $\P_0$, for different values of Reynolds number $\Reyn_0\in[10^3,10^4]$.} 
\label{fig:maxP_FT}
\end{center}
\end{figure}

On the other hand, it follows from estimate \eqref{eq:maxP_timeDep} that
\begin{equation}\label{eq:Qmax}
Q_{\max} :=  4\nu\frac{\P^{1/2}(t_{\max}) - \P^{1/2}(0)}{\E(0) - \E(t_{\max})} \leq
a + b\sqrt{\ln \Reyn_0 + c}.
\end{equation}
Figure \ref{fig:maxP_FT}(a) shows the dependence of $Q_{\max}$ on $\Reyn_0$, for values of palinstrophy in the range $\P_0 \in [10^6,10^9]$. The curve $\gamma(\Reyn_0) = \tilde{a} +\tilde{b}\sqrt{\ln\Reyn_0 + \tilde{c}}$, with $\tilde{a}$, $\tilde{b}$ and $\tilde{c}$ given in \eqref{eq:Prefactor_Fitted}, is included for reference as a red dashed curve. It can observed from the figure that although estimate \eqref{eq:maxP_timeDep} is saturated only over a short time interval, the dependence of the finite-time growth of palinstrophy on $\Reyn_0$, when compensated by the enstrophy dissipation occurring at the maximum palinstrophy time, has a  behavior similar to the one predicted by estimate \eqref{eq:Qmax}.

Figure \ref{fig:maxP_FT}(b) shows the dependence on $\P(0)$ of the compensated maximum palinstrophy
\begin{equation}
\P_{\max} = \P(0)^{-1}\,\mathop{\max}_{t > 0} \, \P(t)
\end{equation}
obtained from the time evolution of the optimal vorticity $\twRP$, for fixed $\Reyn_0$.
As expected from estimate \eqref{eq:maxP_aPriori}, the data indicates that $\P_{\max}$ is independent of $\P(0)$, with its value depending only on the initial Reynolds number $\Reyn_0$.
These results provide compelling evidence supporting the sharpness of the \emph{a priori} finite-time estimate \eqref{eq:maxP_aPriori} with respect to $\P(0)$.

\section{Discussion and Conclusion}
\label{sec:Conclusion}

We have presented numerical confirmation that the rigorous analytic estimate
\[
\frac{d\P}{dt} \leq \left(a + b\sqrt{\ln \Reyn + c}\right)\P^{3/2}
\]
is sharp in its behavior with respect to both the palinstrophy $\P$ and the Reynolds number $\Reyn$.
The power-law dependence on $\P$ is predicted by the self-similar analysis from Appendix \ref{sec:SelfSimilarAnalysis} and confirmed by the data shown in Figure \ref{fig:maxdP_vsP}, where the compensated optimal rate of growth of palinstrophy $\P_0^{-3/2}\R(\tuvecRP)$ is plotted against $\P_0$.
Similarly, the dependence of $\R(\tuvecRP)$ on $\Reyn_0$ is correctly captured by the estimate although the constants $\tilde{a}, \tilde{b}$ and $\tilde{c}$ that fit the optimal rate of growth in the least-squares sense differ from the analytic values given in appendix \ref{sec:dPdt_Estimate}.
More careful analysis might give better constants but it is important to note that the approach used to construct the optimal fields only ensures that solutions to \eqref{eq:OptProb_K0P0} are {\it local} maximizers: there is no guarantee that they are global maximizers of $\R$.
However, the best one can hope for in the search of any other maximizers is to improve the value of the constants $\tilde{a}$, $\tilde{b}$ and $\tilde{c}$ so that optimal instantaneous production of palinstrophy matches that given by the analytic estimate. 

Regarding the finite-time growth of palinstrophy, we have provided evidence supporting the sharpness of the \emph{a priori} estimate
\[
\mathop{\max}_{t > 0} \, \P(t) \leq \psi(\Reyn_0) \P(0)
\]
with respect to the initial palinstrophy $\P(0)$. For this, we have used the instantaneously optimal fields $\tuvecRP$ as initial condition in the 2-D incompressible Navier-Stokes equation, and carefully monitored the time evolution of palinstrophy. The sharpness with respect to the prefactor $\psi(\Reyn_0)$, on the other hand, is a more subtle issue and instead we looked at the point-wise estimate
\[
\P^{1/2}(t) - \P^{1/2}(0) \leq \frac{\gamma(\Reyn_0)}{4\nu} \left( \E(t) - \E(0) \right).
\]
This time-dependent estimate was found to be saturated by the instantaneously optimal fields 
$\tuvecRP$ only over a short time window, resulting in a sub-optimal dependence of the compensated maximum growth of palinstrophy on $\Reyn_0$. This does not mean, however, that the estimate is not sharp, as the fields $\tuvecRP$ are optimal only at time $t=0$. A better assessment of the sharpness of the point-wise estimate \eqref{eq:maxP_timeDep} could be performed by solving the finite-time optimization problem:
\begin{equation}\label{eq:OptProb_FT}
\begin{aligned}
 & \qquad\qquad\qquad\mathop{\max}_{\uvec\in\mathcal{S}} \;\; \P(\uvec(\cdot,T)), 
\quad {\textrm{with}}\\
\mathcal{S} = & \{ \uvec\in H^2(\Omega) : \nabla\cdot\uvec = 0, \;  \K(\uvec(\cdot,0)) =\K_0, \; \P(\uvec(\cdot,0)) = \P_0, \uvec\mbox{ solves (1.1)} \},
\end{aligned}
\end{equation}
with the constraint manifold $\mathcal{S}$ including only the fields which are solutions to the 2-D incompressible Navier-Stokes equation defined on the interval $(0,T)$. This study is, however, outside of the scope of this manuscript and it is left as an open question for subsequent work.

A possible interpretation of the results presented in this manuscript is that it is possible to saturate \emph{a priori} finite-time estimates using fields that saturate instantaneous estimates. Time-dependent point-wise estimates, on the other hand, are saturated only over short time intervals, rendering the finite time growth of palinstrophy obtained from instantaneously optimal fields suboptimal. This implies that in the context of 3-D Navier-Stokes equation, for which no \emph{a priori} finite-time estimates for the growth of enstrophy are available and only time-dependent point-wise estimates exist, the search of solenoidal fields that are optimal for the growth of enstrophy, the key quantity controlling the regularity of solutions in 3-D, must be performed following a finite-time optimization approach where the Navier-Stokes equation is included as part of the constraint, similar to the approach described in equation \eqref{eq:OptProb_FT}. Although computationally intensive, this task is within reach with current available resources and it is also left as an open question for subsequent work. 

\section*{Acknowledgements}

The authors are extremely grateful to Felix Otto for the useful suggestions and discussions leading to the results presented in this manuscript.
This research was supported in part by a NSF Award PHY-1205219, a 2014 Simons Foundation Fellowship in Theoretical Physics, NSF Award DMS-1515161, and a 2016 John Simon Guggenheim Foundation Fellowship in Applied Mathematics.

\appendix

\section{Estimates for the growth of palinstrophy}
\label{sec:Analytic_Estimates}

\subsection{Instantaneous growth of palinstrophy}
\label{sec:dPdt_Estimate}

From equation \eqref{eq:dQdt_defn}, the instantaneous rate of growth of palinstrophy $d\P/dt$ is defined as:
\begin{equation*}
\frac{d\P}{dt}(\uvec) = -\nu\|\laplacian\omega \|_2^2 - \int_\Omega \nabla\omega\cdot\nabla\uvec\cdot\nabla\omega \;d\Omega,
\end{equation*}
with $\uvec:\Omega\to\mathbb{R}^2$ such that $\nabla\cdot\uvec = 0$, and $\omega = \partial_1 u_2 - \partial_2 u_1$. As the quadratic form induced by $\nabla \uvec$ only depends on its symmetric part $(\nabla \uvec)_{S} := \frac{1}{2}\left(\nabla \uvec + \nabla \uvec^T\right)$ and is bounded by its spectral norm $ | (\nabla \uvec)_{S} |_{\sigma} := \max\{|\lambda_1|,|\lambda_2|\}$, with $\lambda_1$ and $\lambda_2$ being the two real eigenvalues of $(\nabla \uvec)_{S}$, it follows that:
\begin{equation}\label{eq:R_estim_S1}
\begin{aligned}
\frac{d\P}{dt}(\uvec) & \leq -\nu\| \laplacian \omega \|^2_2 + 
\|(\nabla\uvec)_{S}\|_{\sigma,\infty} \int_{\Omega} |\nabla\omega |^2 \; d\Omega \\
 & = -\nu\| \laplacian \omega \|^2_2 + 2\|(\nabla\uvec)_{S}\|_{\sigma,\infty}\P. 
\end{aligned}
\end{equation}
For $\Omega$ a square of side $L$ endowed with periodic boundary conditions, it follows that a function $u:\Omega\to\mathbb{R}$ and its gradient $\nabla u$ admit a Fourier representation of the form:
\begin{equation*}
u(\xvec) = \sum_{\kvec\in \mathbb{Z}_0} \uhat(\kvec)\,e^{ i \kvec\cdot\xvec}, \quad 
\nabla u(\xvec) = \sum_{\kvec\in\mathbb{Z}_0} i \, \kvec \, \uhat(\kvec) \, e^{ i \kvec\cdot\xvec},
\end{equation*}
where $\mathbb{Z}_0 = \left(\frac{2\pi}{L}\right)\left(\mathbb{Z}\times\mathbb{Z}\right)$. In Fourier space, the incompressibility condition $\nabla\cdot\uvec = 0$ takes the form $\kvec \cdot \widehat\uvec(\kvec)=0$ for each $\kvec \in \mathbb{Z}_0 \setminus \{0\}$, which implies that $\widehat\uvec(\kvec) \in \mathbb{C}\kvec^\perp$ with $\kvec^\perp= (-k_2, k_1)$. Together with the fact that the matrix $\frac{1}{2} (\kvec \otimes \kvec^\perp + \kvec^\perp \otimes \kvec)$ has eigenvalues $-\frac{1}{2}|\kvec|^2$ and $\frac{1}{2}|\kvec|^2$ we get the somewhat crude estimate
\begin{equation*}
| (\nabla \uvec)_{S}(\xvec) |_{\sigma} \leq \sum_{\kvec\in\mathbb{Z}_0 \setminus \{0\}} \frac{|\kvec||\hat \uvec(\kvec)|}{|\kvec|^2} \left| \frac{1}{2} (\kvec \otimes \kvec^\perp + \kvec^\perp \otimes \kvec) \right|_{\sigma}  \leq \frac{1}{2}\sum_{\kvec\in\mathbb{Z}_0} | \kvec | | \hat\uvec(\kvec) |
\end{equation*}
Grouping the wave numbers into small, intermediate and high frequencies gives:
\begin{equation*}
| (\nabla \uvec)_{S}(\xvec) |_{\sigma} \leq \, \frac{1}{2}\left(
\sum_{|\kvec| \leq \Lambda_1} | \kvec | | \hat\uvec(\kvec) | \quad + 
\sum_{\Lambda_1 \leq |\kvec| \leq \Lambda_2} | \kvec | | \hat\uvec(\kvec) |  \quad + 
\sum_{|\kvec| \geq \Lambda_2} | \kvec | | \hat\uvec(\kvec) |\right) ,
\end{equation*}
where the cut-off wave numbers $\Lambda_1$ and $\Lambda_2$ are yet to be determined. Each term in the right-hand side of the last inequality can be upper-bounded using the Cauchy-Schwarz inequality as:
\begin{equation*}
\begin{aligned}
\sum_{|\kvec| \leq \Lambda_1} | \kvec | | \hat\uvec(\kvec) |  
 & \leq  \left( \sum_{|\kvec| \leq \Lambda_1} | \kvec |^2 \right)^{1/2} \left( \sum_{|\kvec| \leq \Lambda_1} |\hat\uvec( \kvec ) |^2 \right)^{1/2}\\
 &  \leq \left( 2\pi \int_0^{\Lambda_1}k^3dk \right)^{1/2}\left(2\K\right)^{1/2} \; = \; \sqrt{\pi}\;\Lambda^2_1\;\K^{1/2},
\end{aligned}
\end{equation*}
\begin{equation*}
\begin{aligned}
\sum_{\Lambda_1 \leq |\kvec| \leq \Lambda_2} | \kvec | | \hat\uvec(\kvec) |  
 & \leq \left( \sum_{\Lambda_1 \leq |\kvec| \leq \Lambda_2} \frac{1}{ | \kvec |^2 }  \right)^{1/2}
 \left( \sum_{\Lambda_1 \leq |\kvec| \leq \Lambda_2}  | \kvec | ^4 | \hat\uvec(\kvec) |^2 \right)^{1/2} \\
 & \leq \left(2\pi \int^{\Lambda_2}_{\Lambda_1} \frac{dk}{k}  \right)^{1/2} \left(2\P\right)^{1/2} \; = \; 2\sqrt{\pi}\sqrt{\ln\left(\frac{\Lambda_2}{\Lambda_1} \right)}\; \P^{1/2},
\end{aligned}
\end{equation*}
and
\begin{equation*}
\begin{aligned}
\sum_{|\kvec| \geq \Lambda_2} | \kvec | | \hat\uvec(\kvec) | 
 & \leq  \left( \sum_{ |\kvec| \geq \Lambda_2} \frac{1}{ | \kvec |^4 }  \right)^{1/2}
  \left( \sum_{ |\kvec| \geq \Lambda_2} | \kvec | ^6 | \hat\uvec(\kvec) |^2 \right)^{1/2} \\
 & \leq \left( 2\pi \int^{\infty}_{\Lambda_2} \frac{dk}{k^3}\right)^{1/2} \| \laplacian\omega \|_2 \; = \;
 \frac{\sqrt{\pi}}{\Lambda_2} \| \laplacian \omega \|_2.
\end{aligned}
\end{equation*}
Therefore, the estimate for $\| (\nabla\uvec)_{S} \|_{\sigma,\infty}$ reads:
\begin{equation*}
\| (\nabla\uvec)_{S} \|_{\sigma,\infty} \leq \frac{1}{2}\left(\sqrt{\pi}\;\Lambda^2_1\;\K^{1/2} + 2\sqrt{\pi}\sqrt{\ln\left(\frac{\Lambda_2}{\Lambda_1} \right)}\; \P^{1/2} +  \frac{\sqrt{\pi}}{\Lambda_2} \| \laplacian \omega \|_2\right),
\end{equation*}
with the estimate in \eqref{eq:R_estim_S1} leading to:
\begin{equation*}
\frac{d\P}{dt}(\uvec)  \leq -\nu\| \laplacian \omega \|^2_2 + \sqrt{\pi}\;\Lambda^2_1\;\K^{1/2}\P + 2\sqrt{\pi}\sqrt{\ln\left(\frac{\Lambda_2}{\Lambda_1} \right)}\; \P^{3/2} +  \frac{\sqrt{\pi}}{\Lambda_2} \| \laplacian \omega \|_2\P. 
\end{equation*}
The sum $-\nu\| \laplacian\omega \|_2^2 + 2\sqrt{\pi}\| \laplacian\omega\|_2\P/\Lambda_2$ of the first and last terms in the last inequality is maximal when $\| \laplacian\omega \|_2 = \sqrt{\pi}\P/(2\nu\Lambda_2)$, attaining a maximum value of $\pi\P^2/(4\nu\Lambda^2_2)$. Thus, the estimate for $d\P/dt$ reads:
\begin{equation*}
\begin{aligned}
\frac{d\P}{dt}(\uvec)  & \leq  \frac{\pi}{4}\frac{\P^2}{\nu\Lambda^2_2} + \sqrt{\pi}\;\Lambda^2_1\;\K^{1/2}\P + 2\sqrt{\pi}\sqrt{\ln\left(\frac{\Lambda_2}{\Lambda_1} \right)}\; \P^{3/2} \\
 & \leq \left( \frac{\pi}{4}\frac{\P^{1/2}}{\nu\Lambda^2_2} + 
 \sqrt{\pi}\;\frac{\Lambda^2_1\K^{1/2}}{\P^{1/2}} + 
 2\sqrt{\pi}\sqrt{\ln\left(\frac{\Lambda_2}{\Lambda_1} \right)}\right) \P^{3/2}.
\end{aligned}
\end{equation*}
The cut-off wave numbers  $\Lambda_1$ and $\Lambda_2$ can be chosen so that:
\begin{equation*}
\Lambda_1^2 = a\frac{\P^{1/2}}{\K^{1/2}},\quad\mbox{and}\quad 
\Lambda_2^2 = \frac{1}{b}\frac{\P^{1/2}}{\nu},
\end{equation*}
for $a$ and $b$ positive dimensionless numbers. For this choice of $\Lambda_1$ and $\Lambda_2$, and the introduction of the dimensionless parameter $\Reyn = \K^{1/2}/\nu$, the estimate becomes:
\begin{equation}\label{eq:dPdt_Est_gP32}
\frac{d\P}{dt} \leq \left( \sqrt{\pi} \; a + \frac{\pi}{4} \; b + 
 \sqrt{2\pi}\sqrt{\ln\left(\frac{\mbox{Re}}{ab} \right)} \right)\P^{3/2} = g(a,b)\;\P^{3/2}.
\end{equation}
For sufficiently large $\Reyn$, the prefactor $g(a,b)$ in the power-law estimate from \eqref{eq:dPdt_Est_gP32} is minimized when $\partial g /\partial a = 0$ and $\partial g/\partial b = 0$, i.e.\ the optimal values $(\tilde{a},\tilde{b})$ satisfy:
\begin{equation*}
\tilde{a} = \frac{1}{\sqrt{2}}\left[\ln\left(\frac{\Reyn}{\tilde{a}\tilde{b}} \right)\right]^{-1/2} \quad\textrm{and}\qquad
\tilde{b} = \frac{2\sqrt{2}}{\sqrt{\pi}}\left[\ln\left(\frac{\Reyn}{\tilde{a}\tilde{b}} \right)\right]^{-1/2}.
\end{equation*}
It follows that $\tilde{a} = (\sqrt{\pi}/4)\;\tilde{b}$, with the minimum value $g(\tilde{a},\tilde{b})$ given by:
\begin{equation}\label{eq:Optim_Prefactor}
\mathop{\min}_{(a,b)\in \textrm{QI}} \; g(a,b) = g(\tilde{a},\tilde{b}) = \sqrt{2\pi}\left(\sqrt{|z_1|} + \frac{1}{\sqrt{|z_1|}}\right),
\end{equation}
where QI denotes the first quadrant in the $(a,b)$-plane, and $z_1 = - 1/(2\tilde{a}^2)$ is the smallest of the two solutions of the transcendental equation
\begin{equation}\label{eq:Transcend_Re}
ze^{z} = - \frac{2}{\sqrt{\pi}}\Reyn^{-1}.
\end{equation}
For values of $\Reyn$ satisfying $\Reyn > 2e/\sqrt{\pi}$, the two solutions to equation \eqref{eq:Transcend_Re} are given by
\begin{equation*}
z_0 = W_0\left(\frac{-2}{\sqrt{\pi}\Reyn}\right)\qquad\textrm{and}\qquad
z_1 = W_{-1}\left(\frac{-2}{\sqrt{\pi}\Reyn}\right),
\end{equation*} 
where $W_k$ is the $k$-branch of the Lambert $W$ function. The asymptotic expansion for $W_{-1}$ is given by  \cite{Corless1996}:
\begin{equation*}
W_{-1}(x) \sim \ln(-x) - \ln(-\ln(-x))\quad\textrm{as}\quad x\to 0^-, 
\end{equation*}
from which it follows that, for sufficiently large $\Reyn \geq 2/\sqrt{\pi}$, 
\begin{equation}
z_1 \sim - \left(\ln \Reyn - \ln\left( \frac{2}{\sqrt{\pi}}\right) \right) 
- \ln\left( \ln \Reyn - \ln\left( \frac{2}{\sqrt{\pi}}\right) \right).
\end{equation}
Hence, to leading order in the variable $\ln\Reyn - \ln(2/\sqrt{\pi})$, the optimal prefactor $g(\tilde{a},\tilde{b})$ from equation \eqref{eq:Optim_Prefactor} has the form:
\begin{equation*}
g(\tilde{a},\tilde{b}) = \sqrt{2\pi} \sqrt{\ln\Reyn - \ln\left(\frac{2}{\sqrt{\pi}}\right) },
\end{equation*}
giving the estimate for $d\P/dt$ as:
\begin{equation}\label{eq:dPdt_appendEstimate}
\frac{d\P}{dt} \leq \left(a + b \sqrt{ \ln\Reyn + c } \right) \P^{3/2},
\end{equation}
\[
a = 0, \quad b = \sqrt{2\pi},\quad c = - \ln\left(\frac{2}{\sqrt{\pi}}\right).
\]

\subsection{Finite-time growth of palinstrophy}
\label{sec:maxP_Estimate}

To obtain an \emph{a priori} estimate for the finite-time growth of palinstrophy, we use the energy dissipation equation \eqref{eq:dKdt_defn}, leading to $\K(t) \leq \K(0)$ for all $t > 0$ and, since $\Reyn = \K^{1/2}/\nu$, $\Reyn(t) \leq \Reyn(0) = \Reyn_0$. It should be noted that estimate \eqref{eq:dPdt_appendEstimate} holds only for values of Reynolds $\Reyn \geq 2/\sqrt{\pi}$, thus we expect the estimate to be valid only on the time interval where this constraint is satisfied. Using time integration of \eqref{eq:dPdt_appendEstimate} gives:
\begin{align}
\frac{d\P}{dt} & \leq \left(a + b\sqrt{\ln\Reyn_0 + c}\right)\P^{3/2} \quad \Rightarrow \nonumber \\
\int_{\P(0)}^{\P(t)} \P^{-1/2} d\P & \leq \left( a + b\sqrt{\ln\Reyn_0+c} \right)\int_{0}^t\P(s) \,ds \quad \Rightarrow \nonumber\\
\P^{1/2}(t) - \P^{1/2}(0) & \leq \left( a + b\sqrt{\ln\Reyn_0+c} \right)\left(\frac{\E(0) - \E(t)}{4\nu}\right) \label{eq:maxP_timeDepAppendix}
\end{align} 
where time integration of the enstrophy dissipation equation \eqref{eq:dEdt_defn} has been used. Although \eqref{eq:maxP_timeDepAppendix} is not an \emph{a priori} estimate, i.e. given exclusively in terms of the initial data, it can still be used to determine to what extent sharp instantaneous estimates lead to sharp finite-time estimates. To obtain an \emph{a priori} estimate for the maximum finite-time growth of palinstrophy one can use  the fact that $\E(t) \leq \E(0)$ for all $t>0$, which follows from the enstrophy dissipation equation \eqref{eq:dEdt_defn}, along with the estimate $\E \leq \K^{1/2}\P^{1/2}$ leading to
\[
\P^{1/2}(t) \leq \left( 1 + \frac{a + b\sqrt{\ln\Reyn_0+c}}{4} \,\Reyn_0 \right) \, \P^{1/2}(0), 
\]
which gives the \emph{a priori} finite-time estimate
\begin{equation}\label{eq:finiteTime_appendEstimate}
\mathop{\max}_{t\geq0} \, \P(t) \leq  C_{\Reyn_0} \P(0) \quad\mbox{with}\quad 
C_{\Reyn_0} = \left(1+ \frac{a + b\sqrt{\ln\Reyn_0+c}}{4} \,\Reyn_0\right)^2,
\end{equation}
where the prefactor $C_{\Reyn_0}$ depends exclusively on the initial Reynolds number $\Reyn_0 = \K^{1/2}(0)/\nu$, and 
\[
a = 0,\quad b = \sqrt{2\pi},\quad c = -\ln\left(\frac{2}{\sqrt{\pi}}\right).
\]

\section{Self-similar Optimal Fields}
\label{sec:SelfSimilarAnalysis}

As discussed in section \ref{sec:2D_InstOpt}, we are interested in finding incompressible fields maximizing the instantaneous production of palinstrophy $d\P/dt$. That is, we are solving the optimization problem
\begin{equation}\label{eq:OptimProb_Vort}
\begin{aligned}
 & \qquad\mathop{\max}_{\omega\in\mathcal{S}_{\K_0,\P_0}} \;\; \R(\omega), 
 \quad {\textrm{with}}\\
\mathcal{S}_{\K_0,\P_0} = & \{ \omega\in H^2(\Omega) :  \;  \K(\omega) = \K_0, \; \P(\omega) = \P_0 \}.
\end{aligned}
\end{equation}
Optimization problem \eqref{eq:OptimProb_Vort} is equivalent to problem \eqref{eq:OptProb_K0P0}, except that it has been written in terms of the vorticity $\omega$ instead of the velocity field $\uvec$. The energy, palinstrophy and objective functionals are given in terms of $\omega$ as:
\begin{subequations}
\begin{align}
\K(\omega) & = \frac{1}{2}\int_\Omega | \nablaperp\psi |^2 \, d\Omega, \\
\P(\omega) & = \frac{1}{2}\int_\Omega | \nablaperp\omega |^2 \, d\Omega,\\
\R(\omega) & = \int_{\Omega}\J(\omega,\psi)\laplacian\omega \, d\Omega - 
\nu \int_{\Omega}\left(\laplacian \omega \right)^2 \, d\Omega,
\end{align}
\end{subequations}    
where $\nablaperp = [\partial_{x_2}, -\partial_{x_1}]^T$,  $\J(f,g) = (\partial_{x_1}f)( \partial_{x_2} g) - (\partial_{x_2} f)( \partial_{x_1} g)$ is the Jacobian determinant, and the streamfunction $\psi$ and the vorticity $\omega$ satisfy the state equation:
\begin{equation}\label{eq:vort2Stream}
-\laplacian\psi = \omega\qquad\mbox{in } \Omega.
\end{equation}

The first-order optimality condition, i.e. the Euler-Lagrange equation for problem \eqref{eq:OptimProb_Vort} reads:
\begin{equation}\label{eq:KKT_Vort}
\frac{\delta\R}{\delta\omega} - \lambda_{\K}\frac{\delta\K}{\delta\omega} - \lambda_{\P}\frac{\delta\P}{\delta\omega} = 0,
\end{equation}
where $\lambda_{\K}$ and $\lambda_{\P}$ are the Lagrange multipliers associated with the constraints defining the manifold $\mathcal{S}_{\K_0,\P_0}$, and the corresponding variations of the functionals $\R$, $\K$ and $\P$ with respect to variations in $\omega$ are given by:
\[
\frac{\delta\R}{\delta\omega} = \J(\psi,\laplacian\omega) + \laplacian\J(\omega,\psi) + \psi^* - 2\nu\laplacian^2\omega, 
\]
\[
\frac{\delta\K}{\delta\omega} = -\psi, \quad\mbox{and}\quad 
\frac{\delta\P}{\delta\omega} = \laplacian\omega, 
\]
with $\psi^*$ representing the Lagrange multiplier associated with the state equation \eqref{eq:vort2Stream} and obtained as the solution to the elliptic problem:
\begin{equation}\label{eq:vort2Stream_Adj}
-\laplacian\psi^* = \J(\laplacian\omega,\omega).
\end{equation}
Evidence supporting the existence of optimal vorticity fields $\twKP$ which, for fixed energy $\K_0$, vary in a self-similar manner with $\P_0$ was presented by \cite{ap13a}. With this numerical evidence in mind, we look for fields $\omega_{\P_0}$, $\psi_{\P_0}$ and $\psi^*_{\P_0}$, satisfying equation \eqref{eq:KKT_Vort} and subject to the constraints $\K(\omega_{\P_0}) = \K_0$ and $\P(\omega_{\P_0}) = \P_0$, of the form:
\[
\begin{aligned}
\omega_{\P_0}(\xvec) & = \P_0^\alpha\Phi(\P_0^q \xvec), \\
\psi_{\P_0}(\xvec) & = \P_0^\beta\Psi(\P_0^q \xvec), \\
\psi^*_{\P_0}(\xvec) & = \P_0^\gamma\Psi^*(\P_0^q \xvec),
\end{aligned}
\] 
for some real parameters $\alpha,\beta,\gamma$ and $q$, and some functions $\Phi$, $\Psi$ and $\Psi^*$  independent of $\P_0$. Using these ansatz and the rescaled spatial variables $\yvec =  \P_0^q\xvec$, it follows that:
\begin{subequations}
\begin{align}
\K(\omega_{\P_0}) & = \P_0^{2\beta}\left( \frac{1}{2}\int | \nablaperp\Psi |^2 \, d\yvec\right) \quad\mbox{and}\\
\P(\omega_{\P_0}) & = \P_0^{2\alpha}\left( \frac{1}{2}\int | \nablaperp\Phi |^2 \, d\yvec\right).
\end{align}
\end{subequations}
From the energy and palinstrophy constraints it follows that $\alpha = 1/2$ and $\beta = 0$, and from state equations \eqref{eq:vort2Stream} and \eqref{eq:vort2Stream_Adj} we obtain $q = 1/4$ and $\gamma = 3/2$. Finally, the Euler-Lagrange equation \eqref{eq:KKT_Vort} reads:
\[
\P_0^{3/2} \left( \J(\Psi,\laplacian\Phi) + \laplacian\J(\Phi,\Psi) + \Psi^* - 
2\nu\laplacian^2\Phi + \lambda_1 \Psi - \lambda_2 \laplacian\Phi \right) = 0, 
\] 
where $\lambda_1$ and $\lambda_2$ are the Lagrange multipliers corresponding to the constraints 
\[
\frac{1}{2}\int | \nablaperp\Psi |^2 \, d\yvec = \K_0 \quad\mbox{and}\quad
\frac{1}{2}\int | \nablaperp\Phi |^2 \, d\yvec = 1.
\]
The objective functional can be thus evaluated for $\omega_{\P_0}$, yielding:
\[
\R(\omega_{\P_0}) = \left( -\nu \int \left( \laplacian\Phi\right)^2\,d\yvec + 
\int \J(\Phi,\Psi)\laplacian\Phi \,d\yvec \right)\P_0^{3/2},
\]
in agreement with the observed optimal instantaneous growth, and in line with the power-law behavior predicted by estimate \eqref{eq:dPdt_LogEstimate}. 
The dependence of the prefactor
\[
C_{\K_0,\nu} = -\nu \int \left( \laplacian\Phi\right)^2\,d\yvec + 
\int \J(\Phi,\Psi)\laplacian\Phi \,d\yvec 
\]
on the ratio $\K_0^{1/2}/\nu$ is the main subject of study of the present work.

\bibliographystyle{alpha}

\end{document}